\documentclass[aps,pra,twocolumn,superscriptaddress,longbibliography]{revtex4-2}

\usepackage{graphicx}
\usepackage{dcolumn} 
\usepackage{bm}      
\usepackage{amsmath}
\usepackage{amssymb}
\usepackage{epstopdf}
\usepackage{xcolor}
\usepackage{comment}
\usepackage[normalem]{ulem}

\graphicspath{{./images/}}

\begin{document}

\title{Charged Bose polarons at finite momentum}

\author{Grover Andrade Sánchez}
\affiliation{Instituto de F\'isica, Universidad Nacional Aut\'onoma de M\'exico, Apartado Postal 20-364, Ciudad de M\'exico C.P. 01000, Mexico}

\author{Arturo Camacho Guardian}
\affiliation{Instituto de F\'isica, Universidad Nacional Aut\'onoma de M\'exico, Apartado Postal 20-364, Ciudad de M\'exico C.P. 01000, Mexico}

\begin{abstract}
Charged impurities in quantum fluids have unveiled new classes of strongly correlated many-body states across condensed matter, ultracold gases, and hybrid atom–ion platforms. While previous studies have primarily focused on their ground-state and static properties, much less is known about their finite-momentum behavior, which governs transport, dissipation, and quasiparticle stability. Here, we investigate the momentum-dependent properties of a charged Bose polaron using a diagrammatic approach within second-order perturbation theory, explicitly accounting for the finite-range nature of the ion–atom interaction. We show that the interaction range introduces a characteristic momentum scale at which many-body dressing and dissipation are maximized, leading to a non-monotonic behaviour of the damping rate and quasiparticle energy. In the high-momentum regime, we uncover a scaling law $\Gamma_p \sim 1/p$, signaling the suppression of many-body dressing and the recovery of quasi-free impurity dynamics, in stark contrast to the divergent behavior predicted by contact-interaction perturbative treatments.
\end{abstract}

\maketitle

\maketitle
\section{Introduction}
The dynamics of impurities in quantum fluids have provided a paradigmatic route to understand dissipation, collective excitations, and emergent quasiparticles. In analogy with Cherenkov radiation in electrodynamics~\cite{cerenkov1934visible}, a mobile impurity traveling in a superfluid can dissipate energy only above a critical momentum~\cite{nielsen2019critical}. In Fermi systems, the dynamics can be richer, as the interplay between impurity recoil and collective excitations produces quantum flutter~\cite{mathy2012quantum,Knap2014,dolgirev2021emergence}, or induce roton-like dispersions~\cite{Schmidt2012,Cotlet2020}, whereas in quantum Bose gases, supersonic impurities in strongly correlated quantum liquids have been shown to generate nontrivial nonequilibrium dynamics analogous to Cherenkov emission~\cite{Seetharam2021,Seetharam2024}, revealing that the motion of fast impurities can fundamentally reorganize the surrounding quantum medium. 

For polarons, then, finite momentum is not solely a kinematic parameter of the impurity, but rather a central ingredient controlling the structure, stability, and dynamical response of emergent quasiparticles. Quantum gases have provided a versatile platform to investigate impurity physics in regimes inaccessible in solids~\cite{massignan2025polarons,grusdt2025impurities}. This has enabled the exploration of Bose polarons in the strongly interacting regime~\cite{hu2016bose,jorgensen2016observation,ardila2015impurity}, their quantum dynamics~\cite{skou2021non,Skou2022}, criticality~\cite{yan2020bose}, and connections to Efimov physics~\cite{Sun2017,christianen2022bose,levinsen2015impurity,Naidon2018}. It has also uncovered new directions involving metastable polarons~\cite{Morgen2025}, mediated interactions~\cite{Naidon2018,paredes2024perspective,levinsen2025role, Levinsen2025a,Morgen2026}, lattice polarons~\cite{ding2023polarons,santiago2024lattice,Santiago-Garcia_2023,Alhyde2022,vashisht2025chiral,alhyder2025lattice,Isaule2025,yordanov2023mobile,Hartweg2025,rojo2024few}, Rydberg and dipolar polarons~\cite{dominguez2026polarons,Schmidt2018,Ardila2018,Artem2023}. 

In parallel, hybrid ion-atom experiments have evolved toward controlled chemistry~\cite{Hirzler2022,Rianne2022}, molecular formation~\cite{Mohammadi2021}, and atom--ion collisions and interactions~\cite{Dieterle2020,Dieterle2021a,Weckesser2021}. The ability to embed and manipulate single ions in complex many-body environments has opened new possibilities for studying charged polarons and exotic correlated states~\cite{chowdhury2024ion,Pessoa2024,Christensen2021,Astrakharchik2021,Astrakharchik2022,Cavazos2024,Simons2024,Wysocki_2026,Cote2002,Massignan2005,Yogurt2025}, as well as mediated interactions~\cite{Ding2022}. Moreover, external fields provide an additional handle to control their dynamics and transport properties~\cite{Wysocki_2026}. Despite this rapid progress, most studies of charged Bose polarons have focused on their static and low-momentum properties. As a result, the finite-momentum regime, where recoil, dissipation, and many-body dressing become intrinsically intertwined, remains largely unexplored, particularly in the presence of long-range ion-atom interactions beyond the contact interaction paradigm.

\begin{figure}[t]
    \centering
    \includegraphics[width=0.92\linewidth]{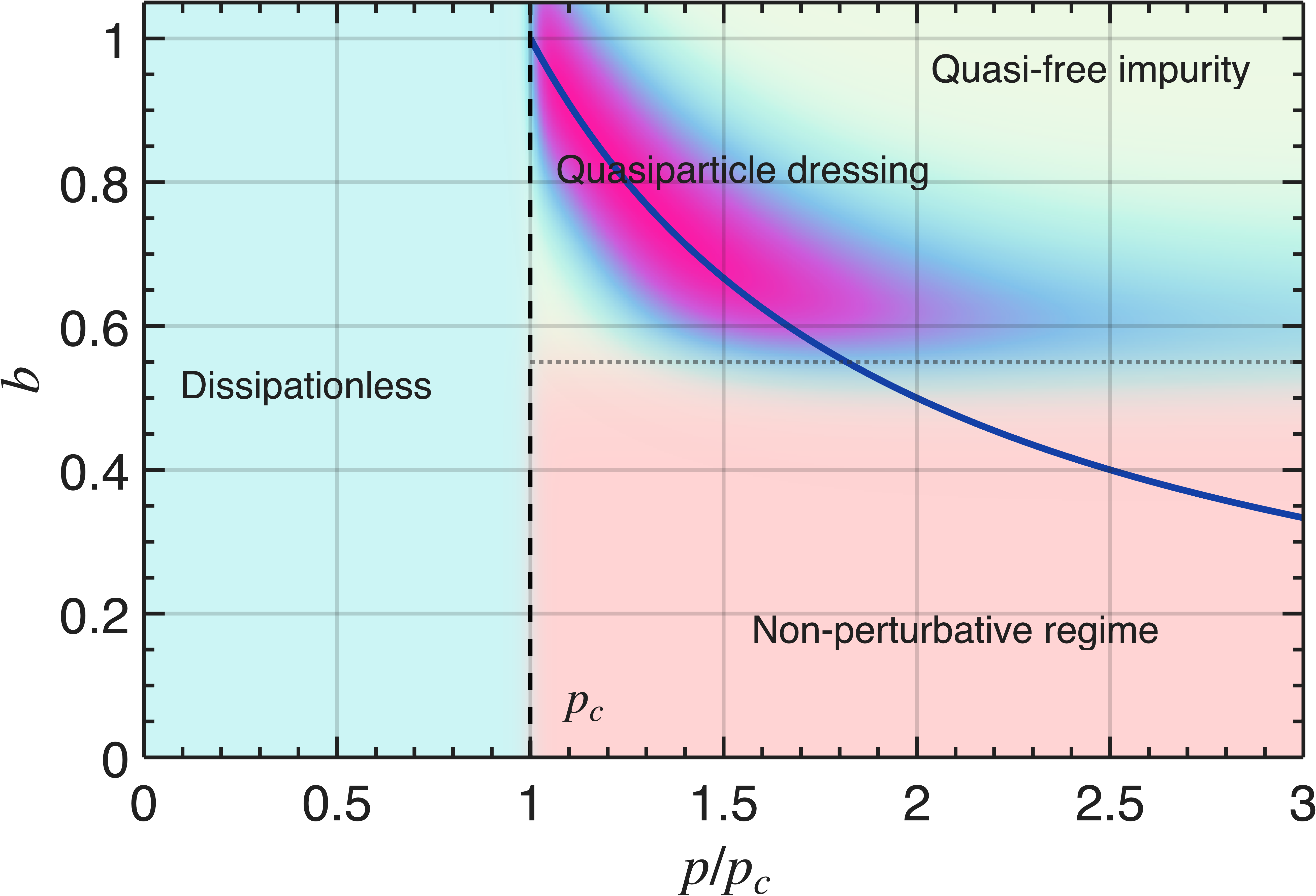}
    \caption{
    Schematic diagram of the finite-momentum charged Bose polaron for the ion-atom interaction.  The vertical dashed line marks the Landau threshold.
    Above this threshold, quasiparticle dressing is strongest around the characteristic scale $p\sim 1/b$.
    For larger momenta or larger effective ranges, the impurity enters a weakly dressed, quasi-free regime where the damping is suppressed.
    The red shadow indicates the non-perturbative regime associated with two-body bound-state formation.  The blue curve marks $p=1/b.$
   }
    \label{fig:ion_atom_phase_diagram}
\end{figure}

Finite-momentum Bose polarons with contact interactions have been studied comprehensively using second-order perturbation theory~\cite{Nielsen2019} and for strong interactions using a coherent ansatz~\cite{Seetharam2021,Seetharam2024}.To the best of our knowledge, the dynamics of charged polarons have only recently been studied 
in Ref.~\cite{Wysocki_2026}, where a mean-field approach in the co-moving frame was used to 
characterize the real-time momentum relaxation and effective mass of the ion. In contrast, our 
diagrammatic treatment provides the momentum-resolved quasiparticle properties, damping rate, 
energy shift, and residue, within a perturbative framework that explicitly accounts for the 
finite-range structure of the ion-atom potential. In contrast to contact-interaction models, the realistic ion-atom interaction introduces an intrinsic length scale that qualitatively modifies the momentum dependence of the quasiparticle properties. In particular, we show that the damping rate exhibits a non-monotonic dependence on momentum, reaching a maximum at a characteristic scale set by the interaction range, where many-body dressing becomes most effective. At higher momenta, the damping rate is progressively suppressed and eventually vanishes following a scaling law $\Gamma_p\sim 1/p$, indicating that the ion smoothly approaches the behavior of a free particle. Our results lead to a phase diagram (see Fig.~\ref{fig:ion_atom_phase_diagram}) where the finite-range ion-atom  interaction plays a central role in controlling dissipation and quasiparticle dressing in charged polarons.

Our perturbative approach provides a simple and systematic framework for understanding charged polarons at finite momentum, while establishing a foundation for future studies in the strong-coupling regime, where the interplay between impurity dynamics and the underlying few- and many-body physics~\cite{Astrakharchik2021,Astrakharchik2022,Christensen2021} may give rise to new emergent regimes.

\section{System}
We consider a single ion coupled to a Bose-Einstein condensate. The Hamiltonian of the system is given by
\begin{gather}
    \hat{H} = \sum_{\textbf{k}}\left(\epsilon^{(b)}_{\mathbf k}\hat{b}^{\dagger}_{\mathbf k}\hat{b}_{\mathbf  k}+\epsilon^{(c)}_{\mathbf k}\hat{c}^{\dagger}_{\mathbf  k}\hat{c}_{\mathbf k}\right)
    +\dfrac{g_{BB}}{2V}\sum_{\mathbf k,\mathbf k',\mathbf q}\hat{b}^{\dagger}_{\mathbf k+\mathbf q}\hat{b}^{\dagger}_{\mathbf k'-\mathbf q}\hat{b}_{\mathbf k'}\hat{b}_{\mathbf k}\\ \nonumber+\dfrac{1}{V}\sum_{\mathbf k,\mathbf k',\mathbf q}V_{\text{ion}}(\mathbf q)\hat{b}^{\dagger}_{\mathbf k+\mathbf q}\hat{c}^{\dagger}_{\mathbf k'-\mathbf q}\hat{c}_{\mathbf k'}\hat{b}_{\mathbf k},
\end{gather}
where $\hat{b}^{\dagger}_{\mathbf k}$ creates a boson with momentum $\mathbf k$ and energy $\epsilon^{(b)}_{\mathbf k}=k^2/2m_B$, where $m_B$ is the mass of the bosons. The boson-boson interaction is short-ranged and modelled by a contact interaction characterised by the interaction strength $g_{BB}=4\pi a_{BB}/m_B$ being $a_{BB}$ the boson-boson scattering length.  We assume that the bosons form a Bose-Einstein condensate well-described by Bogoliubov theory. We introduce a typical energy and momentum associated to the BEC as $E_n=k_n^2/2m_B$ and $k_n=(6\pi^2n_B)^{1/3},$ respectively. Here $n_B$ is the density of the condensate and we take $\hbar=1.$ The operator $\hat{c}^{\dagger}_{\mathbf k}$ creates an ion with momentum $\mathbf k$ and energy $\epsilon^{(c)}_{\mathbf k}=k^2/2m_c,$ where $m_c$ is the mass of the ion. 

The ion-atom interaction in momentum space $V_{\text{ion}}(\mathbf q)$ is the Fourier transform of the following potential in real- space ~\cite{Astrakharchik2021,Astrakharchik2022,Christensen2021,Ding2022} 
\begin{equation}
    V(\mathbf {r}) = -\alpha\dfrac{(r^2-c^2)}{r^2+c^2}\dfrac{1}{(r^2+b^2)^2}.
\end{equation}

The potential is characterised by three parameters: $\alpha,$ which is proportional to the polarisability of the atoms, $b$, related to the depth of the potential, and $c$ to the short-range repulsive part of the ion-atom potential. The recent observation of Feshbach resonance in ion-atom systems allows to consider the parameters of the interaction as tunable~\cite{Weckesser2021}. It is also convenient to introduce a characteristic range $R_*=\sqrt{2m_r\alpha},$ which in typical experiments is expected to satisfy $R_*k_n\sim 1.$ On the other hand, the repulsive barrier is characterised by a distance which is orders of magnitude smaller than the typical interparticle distance of the bosons $k_nc\ll 1,$ and thus, the ion-atom interaction is not sensitive to the very short-range details.

\begin{figure}
\includegraphics[width=\columnwidth]{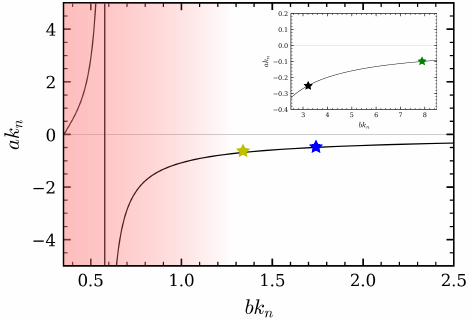}
\caption{Scattering length $ak_n$ of the ion--atom potential as a function of the dimensionless parameter $bk_n$. For sufficiently large $b$, the system lies in the perturbative regime. As $b$ decreases, $ak_n$ exhibits a divergence signaling the onset of a bound state in the two-body problem. The shaded region indicates the regime where bound states emerge and the perturbative description breaks down.}
\label{fig:scattering_length}
\end{figure}

In this work, we will employ a perturbative approach, to this end, we will assume that there is no two-body bound state between the ion and an atom in the BEC. To ensure this, we fix $c=0.0023R_*$ and use values of $b$ larger than $b/R_*=0.58$ where the first molecular state arises. Further details can be found in the Appendix and Refs.~\cite{Astrakharchik2021,Astrakharchik2022,Christensen2021}. Figure~\ref{fig:scattering_length} shows the scattering length associated with the ion-atom interaction as a function of the dimensionless range parameter $bk_n$. In the regime of large $b$, the interaction is weak and the system can be treated perturbatively. As $b$ is reduced, the effective interaction strength increases and the scattering length exhibits a resonance divergence, indicating the formation of a two-body bound state. For smaller values of $b$, the appearance of bound states renders the perturbative treatment invalid. In this work, we restrict our analysis to the weak-coupling regime at large $b$, where no bound states are present. 

\section{Charged polaron: second-order theory}
To characterise the system, we employ a diagrammatic approach using imaginary-time Green's function formalism to develop a perturbative treatment to second order in the ion-boson coupling.  In the context of the ground-state properties of the Bose polaron with contact interactions, a detailed perturbative approach can be found in Ref.~\cite{Christensen2015}. Here, to make the manuscript self-contained, we provide the full derivation.

We introduce the ion Green's function 
\begin{gather}
  \mathcal G_{cc}(\mathbf k,\tau)=-\langle T_{\tau}[\hat c_{\mathbf k}(\tau)\hat c^\dagger_{\mathbf k}(0)] \rangle,
\end{gather}
where $\tau$ denotes the imaginary time and $T_\tau$ is the time-ordering operator. 

In frequency-momentum space, the ion Green's function follows Dyson's equation.
\begin{gather}
\mathcal G^{-1}_{cc}(\mathbf k,z)=z-\epsilon_{\mathbf k}^{(c)}-\Sigma(\mathbf k,z),
\end{gather}
with $z=2\pi i nT$ being the Matsubara frequency, $n\in \mathbb Z$, and $T$ the temperature. Here, the self-energy $\Sigma_{cc}$ incorporates the effects of the bosonic environment on the charged impurity.

In our perturbative treatment, the first-order (mean-field) contribution is
\begin{gather}
 \Sigma_{1}(\mathbf k,z)=E_{\text{MF}}=n_0 V_{\mathrm{ion}}(\mathbf 0)+\sum_{\mathbf p}V_{\mathrm{ion}}(\mathbf 0)n_{b}(E_{\mathbf p})v_{\mathbf p}^2
\end{gather}
where $n_B$ is the boson density and $n_0$ the condensate density. At this point, it is convenient to introduce the BEC Green's function in the Bogoliubov approach. We introduce
\begin{gather}
 \mathcal G_{\text{BEC}}(\mathbf p,z)=\frac{u_\mathbf p^2}{z-E_{\mathbf p}}  -\frac{v_\mathbf p^2}{z+E_{\mathbf p}},\\
 \mathcal F_{\text{BEC}}(\mathbf p,z)=\frac{u_\mathbf pv_{\mathbf p}}{z-E_{\mathbf p}}  -\frac{u_\mathbf pv_{\mathbf p}}{z+E_{\mathbf p}},
\end{gather}
for the normal $\mathcal G_{\text{BEC}}(\mathbf p,z)$ and anomalous $\mathcal F_{\text{BEC}}(\mathbf p,z)$ Green's functions, respectively, written in terms of the coherence factors $u_{\mathbf p}^2-v_{\mathbf p}^2=1$ with $u_{\mathbf p}^2=\frac{1}{2}(\frac{\epsilon_{\mathbf p}+n_0g_{BB}}{E_{\mathbf p}}+1)$ and $E_{\mathbf p}=\sqrt{\epsilon_{\mathbf p}(\epsilon_{\mathbf p}+2g_{BB}n_0)}.$

\begin{figure}[t]
    \centering
    \includegraphics[width=1\linewidth]{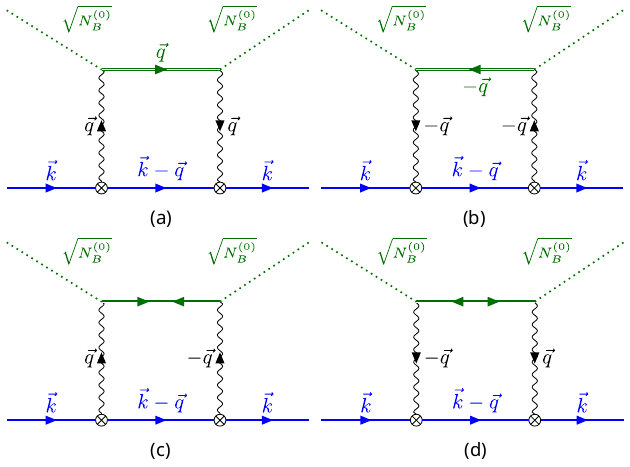}
    \caption{Second-order Feynman diagrams contributing to the self-energy of the charged polaron. 
    Solid blue lines denote the ion propagator, solid green lines represent bosonic propagators, 
    wavy lines indicate the ion-boson interaction, and dashed lines correspond to condensate lines. 
    Panels (a)–(d) account for different processes including normal and anomalous BEC propagators.}
    \label{Fig2}
\end{figure}

Beyond mean-field, the dominant corrections appear at second order in the interaction. The corresponding Feynman diagrams are shown in Fig.~\ref{Fig2}, panels (a)-(d). These diagrams describe virtual processes in which the ion excites a Bogoliubov mode of the condensate and subsequently reabsorbs it, including both normal and anomalous processes. The resulting self-energy contribution reads
\begin{gather}
\Sigma_2(\mathbf p,z) = \dfrac{n_0}{(2\pi)^3}\int \mathrm{d}^3\mathbf q \;
\dfrac{\epsilon^{(b)}_{\mathbf q}}{E_{\mathbf q}}
\dfrac{|V_{\text{ion}}(\mathbf q)|^2}{z-E_{\mathbf q}-\epsilon^{(c)}_{\mathbf p+\mathbf q}}.
\end{gather}

The quasiparticle properties of the charged polaron are determined by the poles of the Green’s function. Expanding around the renormalized energy $\omega_{\mathbf p}$ defined by
\begin{gather}
\omega_{\mathbf p} - \epsilon^{(c)}_{\mathbf p} - \mathrm{Re}\,\Sigma(\mathbf p,\omega_{\mathbf p})=0,
\end{gather}
the impurity propagator acquires the form
\begin{gather}
\mathcal G_{cc}(\mathbf p,z)\simeq \dfrac{Z_{\mathbf p}}{z-\omega_{\mathbf p}+i\Gamma_{\mathbf p}},
\end{gather}
where the quasiparticle residue
\begin{gather}
Z_{\mathbf p}^{-1} = 1 - \left.\dfrac{\partial \mathrm{Re}\,\Sigma(\mathbf p,z)}{\partial z}\right|_{z=\omega_{\mathbf p}}
\end{gather}
measures the overlap between the dressed polaron and the bare ion state. The damping rate, associated with the finite lifetime of the quasiparticle, follows from the imaginary part of the self-energy,
\begin{gather}
\Gamma_{\mathbf p} = - Z_{\mathbf p}\, \mathrm{Im}\,\Sigma(\mathbf p,\omega_{\mathbf p}+i0^+).
\end{gather}
Physically, $\Gamma_{\mathbf p}$ accounts for decay into states where the impurity recoils while exciting Bogoliubov modes. The momentum dependence of $\Gamma_{\mathbf p}$ thus encodes the onset of Landau damping and determines the stability of finite-momentum charged polarons.

In this work, we focus on the damping rate of the charged polaron. Up to second order, the damping rate can also be understood in terms of Fermi's golden rule, 
\begin{gather}
\Gamma_p \;\propto\; n_0\sum_{\mathbf q}\dfrac{\epsilon^{(b)}_{\mathbf q}}{E_{\mathbf q}} |V_{\text{ion}}(\mathbf q)|^2 
\delta\!\left(\epsilon^{(c)}_{\mathbf p}-\epsilon^{(c)}_{\mathbf p-\mathbf q}-E_{\mathbf q}\right).
\label{eq:golden_rule}
\end{gather}
Here, the delta function enforces energy conservation and establishes Landau's superfluid criterion. In this case, energy can only be conserved if $p/m_c > v_c$, where $v_c=\sqrt{n_0g_{BB}/m_B}$ defines the critical velocity of the BEC. That is,  below $v_c$, the impurity cannot emit excitations and moves without friction, while above $v_c$, it undergoes dissipation due to resonant emission of Bogoliubov modes.

\section{Results}
The main focus of our study is the momentum dependence of the quasiparticle properties. In particular, we analyze the damping rate and quasiparticle energy of the charged polaron. Details of the calculation of the self-energy, spectral function, and quasiparticle residue are provided in the Appendix.

\subsection{Damping rate}

In Fig.~\ref{Fig3} we show the damping rate $\Gamma_p$ of the impurity as a function of the dimensionless parameter $p/k_n$. To understand the features of the charged polaron at finite momentum, it is useful to contrast these properties with the {\it conventional polaron} with a boson-impurity interaction modeled by a contact potential. To study the charged polaron and the {\it conventional } polaron on equal footing, we consider the full ion-atom potential at different values of $b$ with a characteristic scattering length as shown in Fig.~\ref{fig:scattering_length}. We employ these same values of the scattering length for the {\it conventional polaron.} In turn, we consider the following values as shown by the stars in Fig.~\ref{fig:scattering_length}: the yellow star corresponds to $bk_n=1.32$, with $ak_n=-0.7$, the blue one indicates  $bk_n=1.71$, with $ak_n=-0.5$. In the inset, we show the scattering length for weaker ion-atom interactions, here, the black star corresponds to $bk_n=3.2$ with $ak_n=-0.25$ and the green one to $bk_n=7.9$ with $ak_n=-0.1$. We consider a weakly interacting Bose gas with $n_0 a_{BB}^3=4.1\times 10^{-4}$.  We take $m=m_c=m_B.$

In the top panel of Fig.~\ref{Fig3} we show the damping rate for the {\it conventional polaron.} The solid lines correspond to the weakly interacting BEC whereas the dashed lines give the ideal BEC ($a_{BB}=0$) and use the same color coding as the stars in Fig.~\ref{Fig3}.  Below the critical momentum  $p<p_c,$ $(p_c=mv_c)$ we find that for the weakly interacting BEC the damping rate strictly vanishes. This is, of course, in agreement with Landau's superfluid criterion. The dashed lines show that in an ideal BEC, dissipation occurs for any finite $p.$ Interestingly, we find that at large momentum, the damping rate increases as $\Gamma_p\sim a^2|p|$ but also increases without bound at large $p.$  This behavior is unphysical and signals the breakdown of the perturbative treatment for a contact interaction. The shaded regions highlight these regimes: dissipationless, dissipative, and an unphysical high-momentum regime where the approximation ceases to be valid for the {\it conventional polaron}.
\begin{figure}[t]
    \centering
    \includegraphics[width=\linewidth]{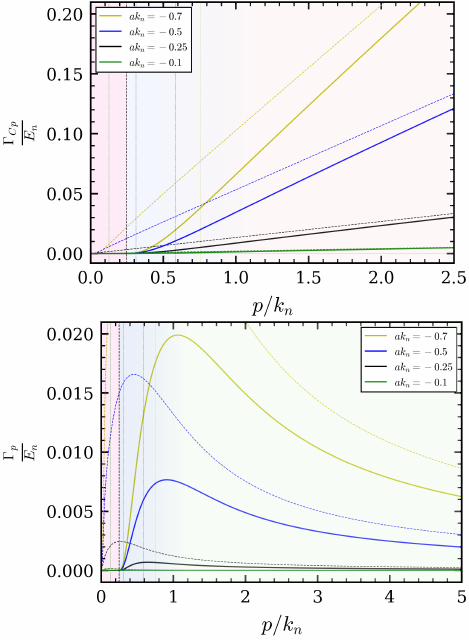}
  \caption{Damping rate $\Gamma_p/E_n$ as a function of momentum $p/k_n$ for (a) contact and (b) ion-atom interactions, shown for equivalent impurity-boson scattering lengths. Solid (dashed) lines correspond to an interacting (ideal) Bose gas. The shaded regions indicate the dissipationless, dissipative, and perturbative breakdown regime (red shadow) or quasi-free regime (green shadow)}
    \label{Fig3}
\end{figure}

Now, we turn to the case of interest, the ion-atom interaction. Here, the damping rate shown in Fig.~\ref{Fig3}(b) exhibits contrasting behavior compared to the {\it conventional polaron}. In the low-momentum regime, we find $\Gamma_p=0$ again in agreement with Landau's criterion for $p<p_c.$ For the ion-atom interaction, the damping rate evolves non-monotonically as a function of momentum above the threshold for damping $p>p_c$. After this onset of dissipation, $\Gamma_p$ increases, reaches a maximum, and subsequently decreases, vanishing in the large-momentum limit. The peak becomes more pronounced and narrower for larger values of the ion-atom scattering length.

The origin of this non-monotonic behaviour can be traced to the length scale $b$ introduced by the ion-atom potential. This scale defines a characteristic momentum $p\sim 1/b$ at which the coupling between the impurity and the Bose gas is most efficient. Around this momentum, the emission of Bogoliubov modes is enhanced, leading to a maximum in the damping rate. For $p\gg 1/b$, we find a strong suppression of the damping. This can be intuitively understood. As the impurity momentum increases, it eventually moves too fast for the bosonic medium to respond. As a result, the dressing cloud cannot form efficiently, and the  impurity asymptotically approaches a bare particle, leading to a suppression of the damping rate. This contrasts sharply with the contact interaction, where no intrinsic length scale exists and the theory artificially predicts an unbounded increase in the damping.

In the high-momentum regime, we find a characteristic power law for the damping rate $\Gamma_p\sim 1/p$. This tendency can be obtained analytically from Eq.~\ref{eq:golden_rule}. In spherical coordinates, and after integrating over the polar angle, we have
\begin{widetext}
\begin{gather}
 \Gamma_p\sim\frac{n_0}{2\pi^2}\int_0^{\infty} dq q^2|V_{\text{ion}}(q)|^2\dfrac{\epsilon^{(b)}_{q}}{E_{ q}}\int_{-1}^{1} dx\, \delta\left(\frac{p q }{m} x-\frac{q^2}{2m}-E_{q} \right) =\frac{n_0\,m}{2\pi^2\,p}\int_0^\infty dq \dfrac{\epsilon^{(b)}_{q}}{E_{ q}}q|V_{\text{ion}}(q)|^2 \Theta(1-|x_0|)
\end{gather}
\end{widetext}
with $x_0$ the solution to the energy-conservation condition: $x_0=
\frac{m}{pq}
\left(
E_q+\frac{q^2}{2m}
\right).$ Then, we obtain the $1/p$ behaviour from the pre-factor of the integral; however, $x_0$ still depends on $p$, which makes the integral itself momentum dependent. In the limit $p\to\infty$, one has $x_0\to 0$ such that $\Theta(1-|x_0|)\to 1$, and for a finite-range interaction the remaining $q$-integral approaches a constant independent of p, leading to the asymptotic scaling $\Gamma_p\sim 1/p$.  That is, the $1/p$ scaling is not a fingerprint of the microscopic interaction. Rather, it reflects the universal kinematics of a fast impurity, whereas the finite-range potential determines only the strength of the residual coupling to the medium.
A detailed derivation based on our diagrammatic approach can be found in the Appendix. 

To demonstrate this scaling,  in Fig.~\ref{FigUniversal}  we plot $p\,\Gamma_p$ as a function of momentum. We find that all of the curves saturate at some constant value at large $p$, demonstrating that the damping rate acquires an asymptotic scaling of $\Gamma_p \sim \frac{1}{p}.$ We find that the specific value at which $p\Gamma_p$ saturates does depend on $b$. However, the scaling occurs for all $b,$ and increases with smaller values of $b$ (larger scattering lengths), which can be calculated as detailed in the Appendix.  Interestingly, we find that this scaling is independent of the specific form of $V_{\text{ion}}(q)$ (as long as the q-integral remains finite).

In contrast to contact interaction, which in perturbation theory diverges as $\Gamma_P\sim a|p|$, the ion-atom potential induces a damping rate with an algebraic decay $1/p$. To heal this divergence with a contact potential, one is forced to employ a non-perturbative treatment; in the simplest case, this can be achieved using an effective $T$-matrix theory~\cite{rath2013field}, which gives the self-energy
\begin{gather}
 \Sigma_\mathcal T(\mathbf p,z)=n_0\frac{4\pi a}{1-ia\sqrt{m}\sqrt{\omega-p^2/4m}}   
\end{gather}
leading to an asymptotic damping rate of 
$ \Gamma_p\sim  \frac{8\pi n_0}{p},$ which also has a $1/p$ power-law form. However, in this case, it is a universal scaling, completely independent of the scattering length $a$. In our case, for the ion-atom interaction, the power law arises even at the level of the perturbative approach.
\begin{figure}[t]
    \centering
    \includegraphics[width=\linewidth]{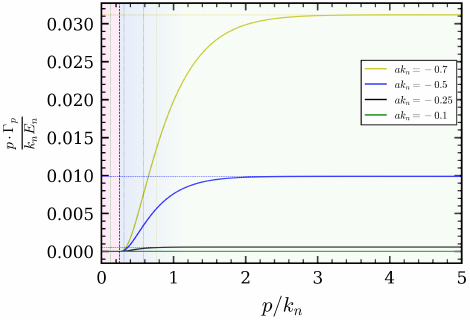}
  \caption{Momentum dependence of the scaled damping rate $p\,\Gamma_p/E_n$ for different interaction strengths $a k_n$. This reveals that all curves approach a constant asymptotic value, demonstrating the asymptotic high-momentum scaling $\Gamma_p \propto \frac{1}{p}.$
}
    \label{FigUniversal}
\end{figure}

\subsection{Quasiparticle Energy}

\begin{figure}[t]
    \centering
    \includegraphics[width=\linewidth]{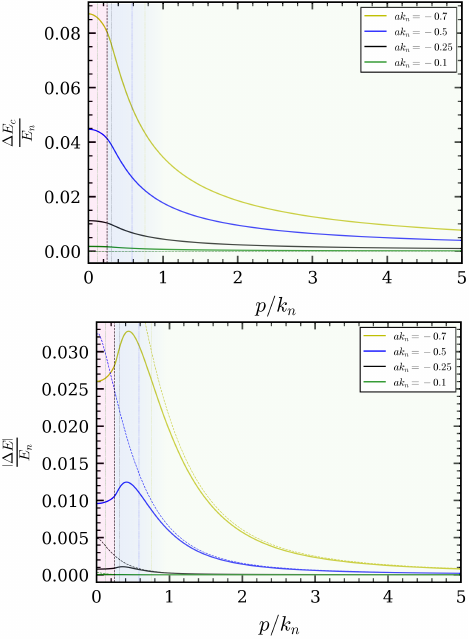}
    \caption{Energy shift $\Delta E/E_n$ as a function of momentum $p/k_n$ for (a) contact and (b) ion-atom interactions, shown for several impurity-boson scattering lengths. Solid (dashed) lines correspond to an interacting (ideal) Bose gas. } 
    \label{Fig5}
\end{figure}

 To obtain a complete picture of the charge polaron, we study the quasiparticle energy.  In Fig.~\ref{Fig5}, we show the interaction-induced energy shift  $\Delta E=|\omega_{\mathbf p} - p^2/2m - E_{\rm MF}|$ of the impurity as a function of momentum. Here, we isolate the effects of the interactions, we subtract both the bare kinetic energy and the mean-field contribution. With this definition, $\Delta E$ directly shows the second-order correction to the quasiparticle energy (after removing the bare kinetic energy and mean-field shift).

For the contact interaction [Fig.~\ref{Fig5}(a)],
we find that the energy shift is maximal at $p=0$ and decreases with increasing momentum. In this case, the quasiparticle energy then tends to the bare impurity energy. Thus, only the imaginary part of the self-energy, that is, the damping, tends to infinity at large $p$. In the case of the ideal BEC (dashed lines), we obtain that $\Delta E$ vanishes identically for all momenta.

In contrast, the ion-atom interaction [Fig.~\ref{Fig5}(b)] again exhibits a non-monotonic dependence on momentum. Starting from small $p$, the energy shift initially increases, reaching a maximum at a finite momentum before decreasing and ultimately vanishing at large $p$. Again, the maximum is not located at zero momentum but occurs around a characteristic scale set by the interaction range, $p \sim 1/b$. The peak of the energy shift becomes more pronounced for larger values of the ion-atom scattering length. We note that, for the energy shift, the position of the peak also depends on the boson-boson scattering length as for $a_{BB}=0$ the maximum of $\Delta E$ is found at $p=0.$  Thus, revealing that polaron dressing exhibits an interplay between the ion-atom and boson-boson interactions.

\section{Conclusions and Outlook}
The study of impurities in quantum fluids has long served as a paradigm for emergent quasiparticles, and recent advances in hybrid atom-ion systems have brought charged impurities to the forefront of experimental and theoretical research. In this context, understanding the dynamical and transport properties of impurities at finite momentum is essential. While most previous works have focused on ground-state properties, the momentum dependence of charged polarons is directly tied to their stability, decay mechanisms, and mobility in a quantum medium. Our work addresses this gap by providing a systematic analysis of finite-momentum quasiparticle properties, highlighting the qualitative role of the interaction range.

Our main findings reveal that the finite-range ion-atom interaction leads to a fundamentally different momentum dependence from the standard contact approximation. In particular, we identify distinct scaling laws for the damping rate: while the perturbative treatment for the contact interaction predicts an unphysical divergence $\Gamma_p \sim |p|$ at large momentum, the ion-atom interaction yields a physically consistent decay $\Gamma_p \sim 1/p$.  Moreover, the damping rate exhibits a non-monotonic behavior with a maximum at a characteristic momentum $p \sim 1/b$, set by the range of the interaction. This scale, absent in contact models, governs both the quasiparticle dressing and the onset of dissipative processes. Similarly, the interaction-induced energy shift displays a non-monotonic dependence, with a maximum at finite momentum and a vanishing asymptotic value, reflecting the suppression of many-body correlations at high momentum. 

Our results open several directions for future research. On the theoretical side, extending the present framework beyond perturbation theory and incorporating strong-coupling effects or non-Markovian dynamics would provide a more complete description of charged polarons. Experimentally, our predictions could be tested in momentum-resolved measurements of ion transport in ultracold gases, either by tracking the impurity dynamics~\cite{skou2021non,Skou2022} or via spectroscopic protocols~\cite{jorgensen2016observation,hu2016bose}. Thus, the characteristic momentum scale set by the finite interaction range, together with the predicted non-monotonic damping and its high-momentum suppression, should be directly accessible in current atom-ion platforms. Thus,  the characteristic scale $1/b$ and the non-monotonic behavior of the damping rate should be directly observable. More broadly, the interplay between finite-range interactions and out-of-equilibrium dynamics may lead to new regimes of impurity physics, including controlled dissipation, nontrivial transport phenomena, and engineered quasiparticles in hybrid quantum systems.

\section*{Acknowledgments}
The authors acknowledge financial support from UNAM DGAPA PAPIIT Grant No. IA101325 and Project SECIHTI No. CBF2023-2024-1765 and PIIF25.

\section*{Data availability statement}
The data that support the findings of this study are openly available in Zenodo at~~\cite{AndradeSanchez2026Dataset}

\begin{widetext}
\appendix
\section{Second-order self-energy, asymptotic power-law, and spectral function}
\subsection{Self-energies}
For a contact interaction between the impurity and the bosons, the self-energy of a zero-momentum impurity has been derived previously at both zero and finite temperature~\cite{Christensen2015,Levinsen2017}. Here, we provide a detailed derivation in order to make our manuscript self-contained.

We explicitly evaluate the four diagrams shown in Fig.~\ref{Fig2}. We begin with diagram (a), which involves the normal Bogoliubov propagator of the BEC,
\begin{gather}
\Sigma^{(a)}(\mathbf{k},i\omega_n)
=
-\frac{n_B^{(0)}}{\beta V}
\sum_{i\omega_s,\mathbf{q}}
V_{\mathrm{ion}}^2(q)\,
\mathcal{G}_{\mathrm{BEC}}(\mathbf{q},i\omega_s)
\,
\mathcal{G}_{c}(\mathbf{k}-\mathbf{q},i\omega_n-i\omega_s,) .
\end{gather}

After performing the Matsubara-frequency summation, we obtain
\begin{gather}
   \Sigma^{(a)}(\mathbf{k},i\omega_n)
    =
    \frac{n_B^{(0)}}{V}
    \sum_{\mathbf{q}}
    V_{\mathrm{ion}}^2(\mathbf q)
    \Bigg\{
    \frac{u_\mathbf q^2}
    {i\omega_n-E_\mathbf q-\varepsilon^{(c)}_{\mathbf{k}-\mathbf{q}}}
    +
    n_B(E_\mathbf q)
    \left(
    \frac{u_\mathbf q^2}
    {i\omega_n-E_\mathbf q-\varepsilon^{(c)}_{\mathbf{k}-\mathbf{q}}}
    +
    \frac{v_\mathbf q^2}
    {i\omega_n+E_\mathbf q-\varepsilon^{(c)}_{\mathbf{k}-\mathbf{q}}}
    \right)
    \nonumber\\
    \pm n_c\!\left(\varepsilon^{(c)}_{\mathbf{k}-\mathbf{q}}\right)
    \left(
    \frac{u_\mathbf q^2}
    {i\omega_n-E_\mathbf q-\varepsilon^{(c)}_{\mathbf{k}-\mathbf{q}}}
    -
    \frac{v_\mathbf q^2}
    {i\omega_n+E_\mathbf q-\varepsilon^{(c)}_{\mathbf{k}-\mathbf{q}}}
    \right)
    \Bigg\},
\end{gather}
with $n_c$ the impurity distribution function which can be either bosonic or fermionic ($\pm$). The second term of the first line accounts for the population of the bosonic Bogoliubov modes $n_B(E_{\mathbf q}).$

Diagram (b) is closely related to diagram (a), the main difference being that the Bogoliubov propagator carries reversed momentum and frequency, namely
$\mathcal{G}_{\rm BEC}(-i\omega_s,-\mathbf q)$.
Physically, while diagram (a) describes the process in which the impurity scatters by creating or absorbing a Bogoliubov excitation with momentum $\mathbf q$, diagram (b) corresponds to the complementary process associated with the opposite branch of the Bogoliubov Green's function.

The origin of this contribution can be understood from the structure of the normal Bogoliubov propagator,
which contains both particle-like and hole-like components. Reversing the frequency and momentum exchanges the relative role of these two contributions,
and therefore leads to a self-energy with the same structure as diagram (a), but with the Bogoliubov coherence factors interchanged, $u_\mathbf q^2 \leftrightarrow v_\mathbf q^2$.

Diagrams (c) and (d) originate from the anomalous Bogoliubov propagator.
As shown in Fig.~\ref{Fig2}, the two diagrams differ only in the direction of the momentum and frequency flowing through the anomalous propagator. Since the anomalous Green's function satisfies the symmetry relation
$
\mathcal F_{\rm BEC}(i\omega_s,\mathbf q)
=
\mathcal F_{\rm BEC}(-i\omega_s,-\mathbf q),$
both diagrams yield identical contributions to the impurity self-energy and can therefore be evaluated together. Their combined contribution is given by
\begin{gather}
    \Sigma^{(c)+(d)}(\mathbf{k},i\omega_n)
    =
    -2\frac{n_B^{(0)}}{\beta V}
    \sum_{i\omega_s,\mathbf q}
    V_{\rm ion}^2(\mathbf q)\,
    \mathcal F_{\rm BEC}(\mathbf q,i\omega_s)\,
    \mathcal G_c(\mathbf{k}-\mathbf q,i\omega_n-i\omega_s).
\end{gather}

Physically, these diagrams describe processes in which the impurity interacts with the coherent particle-hole pairs that constitute the quantum depletion of the condensate. Such processes are absent in a noninteracting Bose gas and emerge solely due to boson-boson interactions. Consequently, the corresponding contribution is proportional to the anomalous coherence factor $u_\mathbf qv_\mathbf q$.  We obtain
\begin{gather}
    \Sigma^{(c)+(d)}(\mathbf{k},i\omega_n)
    =
    -2\frac{n_B^{(0)}}{V}
    \sum_{\mathbf q}
    V_{\rm ion}^2(\mathbf q)
    \Bigg\{
    \frac{u_\mathbf qv_\mathbf q}
    {i\omega_n-E_\mathbf q-\varepsilon^{(c)}_{\mathbf{k}-\mathbf q}}
    +n_B(E_\mathbf q)
    \left(
    \frac{u_\mathbf qv_\mathbf q}
    {i\omega_n-E_\mathbf q-\varepsilon^{(c)}_{\mathbf{k}-\mathbf q}}
    +
    \frac{u_\mathbf qv_\mathbf q}
    {i\omega_n+E_\mathbf q-\varepsilon^{(c)}_{\mathbf{k}-\mathbf q}}
    \right)
    \nonumber\\
    \pm n_c\!\left(\varepsilon^{(c)}_{\mathbf{k}-\mathbf q}\right)
    \left(
    \frac{u_\mathbf qv_\mathbf q}
    {i\omega_n-E_\mathbf q-\varepsilon^{(c)}_{\mathbf{k}-\mathbf q}}
    -
    \frac{u_\mathbf qv_\mathbf q}
    {i\omega_n+E_\mathbf q-\varepsilon^{(c)}_{\mathbf{k}-\mathbf q}}
    \right)
    \Bigg\}.
\end{gather}

The appearance of the product $u_\mathbf qv_\mathbf q$ highlights the fundamentally coherent nature of these processes.  The Bogoliubov coherence factors combine into the compact form
$ u_\mathbf q^2+v_\mathbf q^2-2u_\mathbf qv_\mathbf q
=
(u_\mathbf q-v_\mathbf q)^2
=
\frac{\varepsilon_\mathbf q}{E_\mathbf q},$
which corresponds to the well-known static structure factor of a weakly interacting Bose gas.  Thus, the second-order self energy is given by
\begin{gather}
    \Sigma_{2}(\mathbf{k},i\omega_n)=\frac{n_B^{(0)}}{V}\sum_{\mathbf{q}}V_{\text{ion}}^2(\mathbf{q})\left(u_{\mathbf q}-v_{\mathbf q}\right)^2\left\{\frac{1}{i\omega_n-E_{\mathbf{q}}-\varepsilon_{\mathbf{k}-\mathbf{q}}^{(c)}}+n_B(E_{\mathbf{q}})\left(\frac{1}{i\omega_n-E_{\mathbf{q}}-\varepsilon_{\mathbf{k}-\mathbf{q}}^{(c)}}+\frac{1}{i\omega_n+E_{\mathbf{q}}-\varepsilon_{\mathbf{k}-\mathbf{q}}^{(c)}}\right)-\right.\nonumber\\
    \left.\hphantom{\Sigma^{(a)}(i\omega_n,\mathbf{k})=\frac{n_{B}^{(0)}}{\beta V}-u_{q}^2\frac{1}{i\omega_n-E_{\mathbf{q}}-\varepsilon_{\mathbf{k}-\mathbf{q}}^{(c)}}-n_B(E_{\mathbf{q}\ne0})\left(u_{q}-v_{q}\right)^2}\pm n_c(\varepsilon_{\mathbf{k}-\mathbf{q}}^{(c)})\left(\frac{1}{i\omega_n-E_{\mathbf{q}}-\varepsilon_{\mathbf{k}-\mathbf{q}}^{(c)}}-\frac{1}{i\omega_n+E_{\mathbf{q}}-\varepsilon_{\mathbf{k}-\mathbf{q}}^{(c)}}\right)\right\}
\end{gather}
At zero temperature, $n_B(E_q)=0$. Moreover, in the dilute-impurity limit considered here, the impurity occupation is negligible, so that
$n_c(\varepsilon^{(c)}_{\mathbf{k}-\mathbf{q}})\simeq 0$.
The second-order self-energy therefore reduces to the compact expression
\begin{gather}\label{Ec:Self-Energy_Second-Order_Ion-Atom}
    \Sigma_{2}(\mathbf{k},i\omega_n)
    =
    \frac{n_B^{(0)}}{(2\pi)^3}
    \int d^3q\,
    V_{\mathrm{ion}}^2(\mathbf q)
    \frac{\varepsilon_\mathbf q}{E_\mathbf q}
    \frac{1}
    {i\omega_n-E_\mathbf q-\varepsilon_{\mathbf{k}-\mathbf{q}}^{(c)}} .
\end{gather}
To obtain a closed expression for the self-energy, we evaluate the momentum integral appearing in Eq.~(\ref{Ec:Self-Energy_Second-Order_Ion-Atom}). Owing to the isotropy of the interaction, it is convenient to employ spherical coordinates and choose the external momentum $\mathbf{p}$ along the polar axis. The integral can then be written as

\begin{gather}
\int d^3q\,
V_{\rm ion}^2(\mathbf q)
\frac{\varepsilon_\mathbf q}{E_\mathbf q}
\frac{1}
{i\omega_n-E_\mathbf q-\varepsilon^{(c)}_{\mathbf{p}-\mathbf{q}}}
=
2\pi
\int_0^\infty dq
\int_0^\pi d\theta\,
V_{\rm ion}^2(q)
\frac{\varepsilon_q}{E_q}
\frac{q^2\sin\theta}
{i\omega_n-E_q-\frac{p^2}{2m}-\frac{q^2}{2m}
+\frac{pq}{m}\cos\theta}.
\end{gather}

The angular integral can be performed analytically. Introducing the variable $u=\cos\theta$, one finds

\begin{gather}
\int_0^\pi
\frac{\sin\theta\,d\theta}
{i\omega_n-E_q-\frac{p^2}{2m}-\frac{q^2}{2m}
+\frac{pq}{m}\cos\theta}
=
-\int_1^{-1}
\frac{du}
{i\omega_n-E_q-\frac{p^2}{2m}-\frac{q^2}{2m}
+\frac{pq}{m}u}
=
\frac{m}{pq}
\ln\!\left(
\frac{
i\omega_n-E_q-\frac{(p-q)^2}{2m}
}{
i\omega_n-E_q-\frac{(p+q)^2}{2m}
}
\right).
\end{gather}

Substituting this result into Eq.~(\ref{Ec:Self-Energy_Second-Order_Ion-Atom}), we obtain the second-order impurity self-energy
\begin{gather}\label{Ec:Self-Energy_Ion-Atom}
\Sigma(\mathbf p,i\omega_n)
=
n_B^{(0)}V_{\rm int}(0)
+
\frac{n_B^{(0)}}{(2\pi)^2}
\int_0^\infty dq\,
V_{\rm int}^2(q)
\frac{\varepsilon_q}{E_q}
\frac{mq}{p}
\ln\!\left(
\frac{
i\omega_n-E_q-\frac{(p-q)^2}{2m}
}{
i\omega_n-E_q-\frac{(p+q)^2}{2m}
}
\right).
\end{gather}

\subsection{Decay rate: Large-momentum behavior}

The asymptotic behavior of the quasiparticle decay rate can be extracted directly from the self-energy derived above. After analytic continuation,
$i\omega_n \rightarrow \omega+i\gamma$, with $\gamma$ infinitesimal, the logarithmic term appearing in Eq.~(\ref{Ec:Self-Energy_Ion-Atom}) can be decomposed into its real and imaginary parts as

\begin{gather}
\mathrm{Re}
\left\{
\ln\left(
\frac{\omega+i\gamma-E_q-\frac{(p-q)^2}{2m}}
{\omega+i\gamma-E_q-\frac{(p+q)^2}{2m}}
\right)
\right\}
=
\ln
\left|
\frac{\omega+i\gamma-E_q-\frac{(p-q)^2}{2m}}
{\omega+i\gamma-E_q-\frac{(p+q)^2}{2m}}
\right|,
\end{gather}

\begin{gather}
\mathrm{Im}
\left\{
\ln\left(
\frac{\omega+i\gamma-E_q-\frac{(p-q)^2}{2m}}
{\omega+i\gamma-E_q-\frac{(p+q)^2}{2m}}
\right)
\right\}
=
\mathrm{Arg}
\left(
\omega+i\gamma-E_q-\frac{(p-q)^2}{2m}
\right)
-
\mathrm{Arg}
\left(
\omega+i\gamma-E_q-\frac{(p+q)^2}{2m}
\right).
\end{gather}

To determine the damping rate, we evaluate the self-energy on shell,
$\omega=\varepsilon_p=\frac{p^2}{2m}.$
The real parts of the two arguments then become
\begin{gather}
\varepsilon_p-E_q-\frac{(p+q)^2}{2m}
=
-\frac{pq}{m}-E_q-\frac{q^2}{2m},
\end{gather}

\begin{gather}
\varepsilon_p-E_q-\frac{(p-q)^2}{2m}
=
\frac{pq}{m}-E_q-\frac{q^2}{2m}.
\end{gather}

The first expression is always negative, implying

\begin{gather}
\mathrm{Arg}
\left(
\varepsilon_p+i\gamma-E_q-\frac{(p+q)^2}{2m}
\right)
=
\pi
-
\tan^{-1}
\left(
\frac{\gamma}
{\frac{pq}{m}+E_q+\frac{q^2}{2m}}
\right).
\end{gather}

The second expression changes sign when $\frac{pq}{m} =E_q+\frac{q^2}{2m},
$ which corresponds to the Landau criterion for the onset of dissipation. In the high-momentum regime, $p\gg p_c,$
the quantity
\(
\frac{pq}{m}-E_q-\frac{q^2}{2m}
\)
is positive over the dominant region of integration. Consequently,

\begin{gather}
\mathrm{Arg}
\left(
\varepsilon_p+i\gamma-E_q-\frac{(p-q)^2}{2m}
\right)
=
\tan^{-1}
\left(
\frac{\gamma}
{\frac{pq}{m}-E_q-\frac{q^2}{2m}}
\right).
\end{gather}

Substituting these expressions into Eq.~(\ref{Ec:Self-Energy_Ion-Atom}) yields
\begin{gather}
\mathrm{Im}\,
\Sigma(\mathbf p,\omega+i0^+)
\approx
\frac{n_B^{(0)}}{(2\pi)^2}
\int_0^\infty dq\,
V_{\rm int}^2(q)
\frac{\varepsilon_q}{E_q}
\frac{mq}{p}
\left[
-\pi
+
\frac{2\gamma}{m}
\frac{pq}
{\left(\frac{pq}{m}\right)^2
-
\left(E_q+\frac{q^2}{2m}\right)^2}
\right].
\end{gather}

Defining
\begin{gather}
I_1=\frac{n_B^{(0)}}{4\pi}\int_0^\infty dq\,V_{\rm int}^2(q)\frac{\varepsilon_q}{E_q}\,m q,
\end{gather}
the imaginary part of the self-energy can be written as
\begin{gather}
\mathrm{Im}\,
\Sigma(\mathbf p,\omega+i0^+)
=
-\frac{I_1}{p}
+
\gamma
\frac{n_B^{(0)}}{2\pi^2}
\int_0^\infty dq\,
V_{\rm int}^2(q)
\frac{\varepsilon_q}{E_q}
\frac{q^2}
{\left(\frac{pq}{m}\right)^2
-
\left(E_q+\frac{q^2}{2m}\right)^2}.
\label{ecSI}
\end{gather}

The second contribution vanishes in the limit \(\gamma\rightarrow0^+\), leaving the leading asymptotic behavior
\begin{gather}
\Gamma_p
=
-\mathrm{Im}\,
\Sigma(\mathbf p,\omega+i0^+)
\simeq
\frac{I_1}{p}.
\end{gather}

We therefore conclude that the finite-range ion-atom interaction gives rise to a characteristic algebraic decay of the damping rate,
\begin{gather}
\Gamma_p \propto \frac{1}{p},
\end{gather}
at large momentum. This behavior reflects the progressive suppression of many-body dressing as the impurity kinetic energy becomes much larger than the characteristic energy scales of the condensate. 

Note that if $\gamma$ is not infinitesimal, the second term in Eq.~\ref{ecSI} can contribute significantly to the damping rate. In general, $\gamma$ contains the intrinsic damping rate of the impurity state, which may be finite. Therefore, the polaron decay rate is enhanced both by the intrinsic impurity damping and by interaction-induced many-body processes.

\subsection{Spectral Function}

\label{sec:spectral_function}
 The spectral function is defined as
\begin{equation}
A(\mathbf{p},\omega)
=
-2\,\mathrm{Im}\,
G_{cc}(\mathbf{p},\omega),
\label{eq:spectral_function}
\end{equation}

The spectral function contains both the coherent quasiparticle contribution and the incoherent many-body continuum. In the weak-coupling regime considered in the main text, the spectral response is dominated by a well-defined quasiparticle peak whose position follows the dressed dispersion relation. The width of this peak is controlled by the imaginary part of the self-energy.

Figure~\ref{fig:spectral_function} shows for the charged Bose polaron. Panel~(a) shows  the  momentum-frequency resolved spectral function. A sharp branch is clearly visible, corresponding to the polaron quasiparticle. The finite linewidth originates from interaction-induced decay processes encoded in the imaginary part of the self-energy.
\begin{figure*}[t]
\centering
\includegraphics[width=\textwidth]{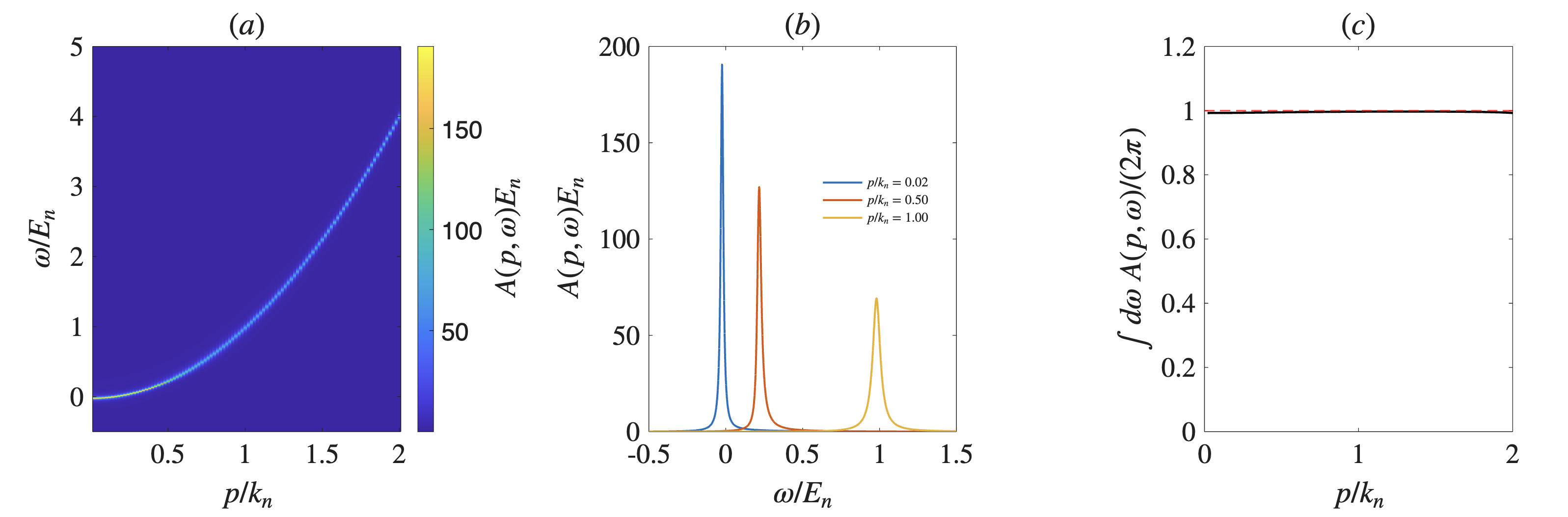}
\caption{
Spectral properties of the charged Bose polaron.
(a) Momentum-resolved spectral function $A(p,\omega)$.
(b) Spectral function at fixed momenta $p/k_n=0.20$, $0.50$, and $1.00$. 
(c) Spectral sum rule,
$\int d\omega\, A(p,\omega)/(2\pi)=1$.}
\label{fig:spectral_function}
\end{figure*}
To illustrate the evolution of the quasiparticle peak, panel~(b) shows representative cuts of the spectral function at fixed momenta. As the momentum increases, the spectral peak shifts to higher energies and broadens slightly, reflecting the increase of the damping rate. Nevertheless, the peak remains narrow, confirming that the quasiparticle remains well defined in this parameter regime.

As a consistency check, panel~(c) verifies the spectral sum rule
\begin{equation}
\int_{-\infty}^{\infty}
\frac{d\omega}{2\pi}
A(\mathbf p,\omega)
=
1.
\label{eq:sumrule}
\end{equation}
The numerical integration remains very close to unity for all momenta. This demonstrates that the computed spectral function satisfies the expected normalization and that the numerical evaluation of the retarded Green's function is well converged.

The fact that the extrema of the damping rate, quasiparticle energy, and residue all occur around the same characteristic scale $p\sim1/b$ demonstrates that the finite interaction range controls the momentum at which many-body dressing is most effective.

\subsection{Quasiparticle Residue}
For completeness, we study the quasiparticle residue. In particular, we focus on two cases: 
(a) the quasiparticle residue as a function of momentum and (b) its dependence on the 
boson-boson scattering length $a_{BB}$. These are shown in Fig.~\ref{FigResidue}.

In Fig.~\ref{FigResidue}(a) we show the quasiparticle residue as a function of momentum for 
several values of the ion-atom scattering length. The residue exhibits a non-monotonic 
dependence on momentum. At $p=0$ the residue satisfies $Z<1$; it then gradually decreases, 
and as the momentum crosses the critical momentum it interestingly becomes $Z>1$, before 
gradually approaching $Z\rightarrow 1$ at large momenta. We stress that, although the spectral 
function satisfies the normalization condition of Eq.~(\ref{eq:sumrule}) and therefore conserves 
the total spectral weight, the quasiparticle residue extracted from the pole approximation may 
become slightly larger than unity when the quasiparticle branch enters the many-body continuum. 
Such values of $Z>1$ should not be interpreted as an excess of spectral weight, but rather as 
an indication that the residue becomes less well defined once the pole is immersed in the 
continuum.

Now, in Fig.~\ref{FigResidue}(right)
The quasiparticle energy and damping rate exhibit a strong dependence on the boson-boson scattering length $a_{BB}$. In the damping rate, changing $a_{BB}$ shifts the onset of Landau damping, whereas in the quasiparticle energy it also displaces the momentum at which the energy shift reaches its maximum. To gain further insight into how boson-boson interactions modify the quasiparticle dressing, we analyze the quasiparticle residue as a function of $a_{BB}$. For clarity, Fig.~\ref{FigResidue}(right) shows the zero-momentum quasiparticle residue $Z$ as a function of the dimensionless interaction strength $a_{BB}k_n$.

We illustrate the quasiparticle residue for both contact interactions (dashed lines) and the ion-atom interaction (solid line) for two different values of the scattering length: $ak_n=-0.7$ (black) and $ak_n=-0.25$ (yellow). We find a strong dependence of the quasiparticle residue for very small values of the boson-boson interaction strength. For large $a_{BB},$ the quasiparticle residue remains significant, thus, leading to a well-defined quasiparticle. However, as $a_{BB}$ is decreased, both interaction potentials lead to the breakdown of the quasiparticle theory as the residue vanishes. This result for the contact potential has already been reported~\cite{Christensen2015,Levinsen2017}.

This indicates that the suppression of the residue in the weakly interacting limit is not an artifact of the contact approximation, but rather a robust feature of the many-body environment treated within second-order perturbation theory. 

\begin{figure}[t]
    \centering
     \includegraphics[width=.4\linewidth]{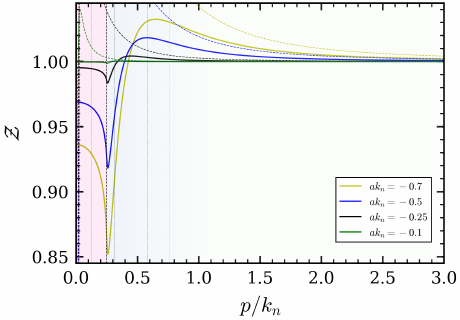}
    \includegraphics[width=.4\linewidth]{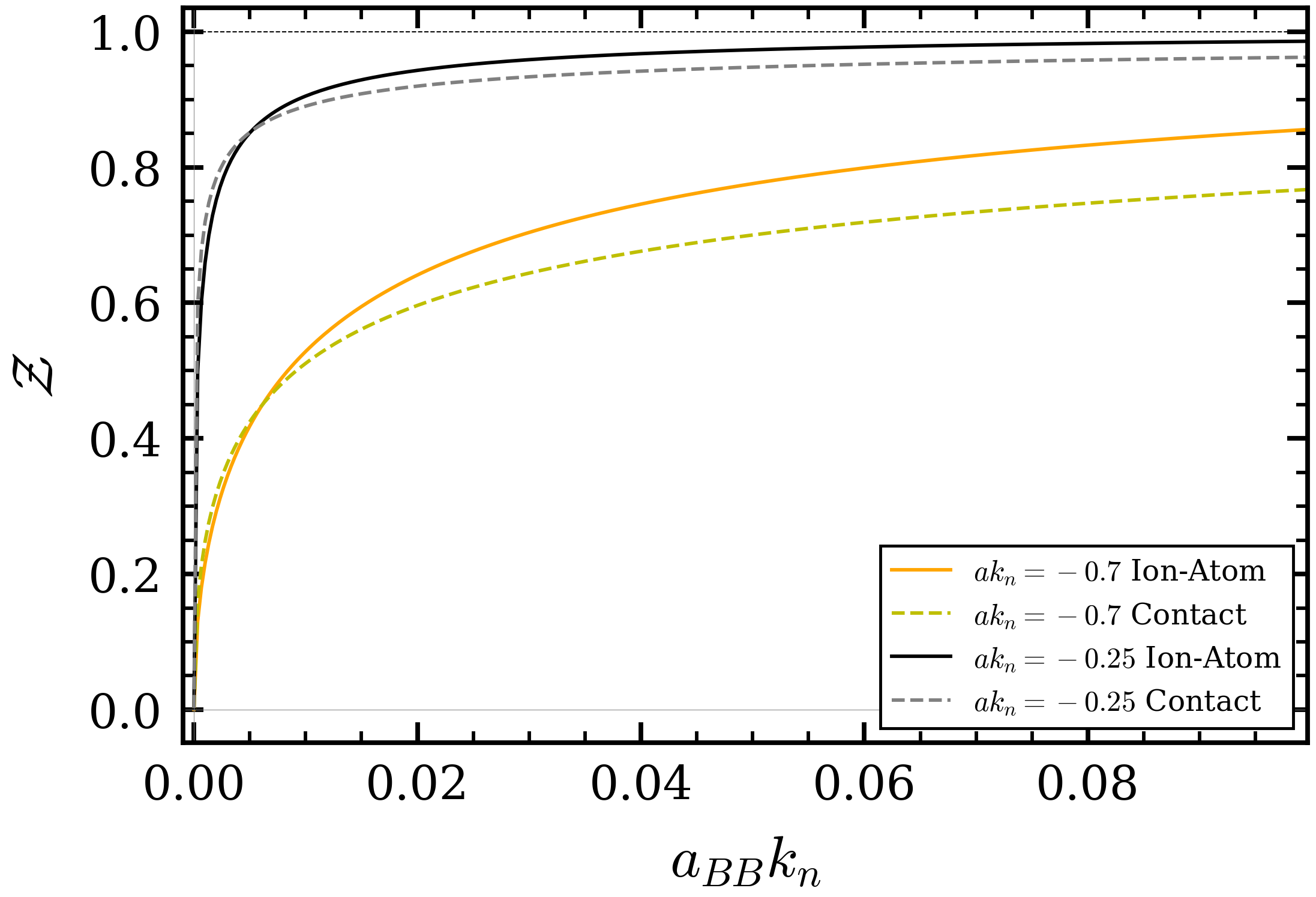}
    \caption{(left)  Quasiparticle residue $Z_p$ as a function of momentum $p/k_n$. (right)Quasiparticle residue $Z$ at zero momentum as a function of the Bose gas interaction strength $a_{BB}k_n$ for contact and ion-atom interactions, shown for two values of the impurity-boson scattering length.}
    \label{FigResidue}
\end{figure}
\end{widetext}

\bibliography{references}
\end{document}